\documentclass[twocolumn,nodate,showpacs,preprintnumbers,amsmath,amssymb,aps,prd]{revtex4}

\usepackage{graphicx}
\usepackage{amsmath,amssymb}

\def\be{\begin{equation}}
\def\ee{\end{equation}}
\def\ba{\begin{eqnarray}}
\def\ea{\end{eqnarray}}

\def\f{\frac}

\def\g{\gamma}

\allowdisplaybreaks

\newcommand{\lp}{\ell_{\mathrm P}}
\newcommand{\ket}[1]{\ensuremath{|#1\rangle}}

\newcommand{\sgn}{\mathop{\mathrm{sgn}}\nolimits}

\newcommand*{\Z}{{\mathbb Z}}

\begin{document}

\title{Effective Scenario of Loop Quantum Cosmology}
%\date{\today}
  \author{You Ding}
    \email{youding@gmail.com}
  \author{Yongge Ma \footnote{Corresponding author}}
     \email{mayg@bnu.edu.cn}
    \author{Jinsong Yang}
    \email{yangksong@gmail.com}
    \affiliation{Department of Physics, Beijing Normal University, Beijing 100875, China}

\begin{abstract}
Semiclassical states in isotropic loop quantum cosmology are
employed to show that the improved dynamics has the correct
classical limit. The effective Hamiltonian for the quantum
cosmological model with a massless scalar field is thus obtained,
which incorporates also the next to leading order quantum
corrections. The possibility that the higher order correction
terms may lead to significant departure from the leading order
effective scenario is revealed. If the semiclassicality of the
model is maintained in the large scale limit, there are great
possibilities for $k=0$ Friedmann expanding universe to undergo a
collapse in the future due to the quantum gravity effect. Thus the
quantum bounce and collapse may contribute a cyclic universe in
the new scenario.
\end{abstract}


\maketitle

The theoretical search for a quantum theory of gravity has been
rather active. The expectation that the singularities predicted by
classical general relativity would be resolved by some quantum
gravity theory has been confirmed by the recent study of certain
isotropic models in loop quantum cosmology (LQC)
\cite{bojowald,analy,improved}, which is a simplified
symmetry-reduced model of a full background independent quantum
theory of gravity \cite{lrr}, known as loop quantum gravity (LQG)
\cite{rev,rov,thie,hhm}. The basis purpose of LQG is to merge the
conceptual insight of general relativity (GR) into quantum
mechanics \cite{rov}. To achieve this purpose, one only makes use
of the general tools of a quantum theory. The Hilbert space and
and operators are obtained from classical GR following certain
quantization strategy. In contrast to the initial Wheeler-DeWitt
canonical quantization of GR \cite{wd}, the classical algebra that
one wants to represent on the Hilbert space of LQG is based on the
holonomies of the gravitational connection. Physically, holonomies
are natural variables representing Faraday's "lines of force",
that do not refer to what happens at a point, but rather refer to
the relation between different points connected by a line.
Mathematically, the quantum configuration space of LQG can be
constructed by the concept of holonomy, since its definition does
not depend on an extra background. Then the kinematical framework
of LQG can be established with mathematical rigour \cite{thie}.

The idea that one should view holonomies rather than connections
as basic variables for the quantization of gravity is successfully
carried on in the models of LQC \cite{first,math,lrr}. In a LQC
scenario for a universe filled with a massless scalar field, the
classical singularity gets replaced by a quantum bounce
\cite{improved,bojow-r}. Various features of the bounce have been
revealed through different considerations
\cite{bojow-r,bojow-prl,cs}. While the model shows that the
quantum effect played the key role in Planck scale to cure the big
bang singularity as one expected, the question whether quantum
gravity effect can also be manifested in large scale cosmology
remains open. This question is crucial since, besides overcoming
the difficulties of a classical theory, to predict phenomena which
are dramatically different from those of the classical theory is
also a hallmark to identify a quantum theory.

The framework that we are considering is the so-called improved
dynamics of LQC \cite{improved}. In the kinematical setting, one
has to introduce an elementary cell ${\cal V}$ and restricts all
integrations to this cell. Fix a fiducial flat metric
${{}^o\!q}_{ab}$ and denote by $V_o$ the volume of ${\cal V}$ in
this geometry. The gravitational phase space variables
---the connections and the density weighted triads
--- can be expressed as $ A_a^i = c\, V_o^{-(1/3)}\,\,
{}^o\!\omega_a^i$ and $E^a_i = p\,
V_o^{-(2/3)}\,\sqrt{{}^o\!q}\,\, {}^o\!e^a_i$, where
$({}^o\!\omega_a^i, {}^o\!e^a_i)$ are a set of orthonormal
co-triads and triads compatible with ${{}^o\!q}_{ab}$ and adapted
to ${\cal V}$. $p$ is related to the scale factor $a$ via
$|p|=V_o^{2/3}a^2$. The fundamental Poisson bracket is given by: $
\{c,\, p\} = {8\pi G\gamma}/{3} $, where $G$ is the Newton's
constant and $\gamma$ the Barbero-Immirzi parameter. The
gravitational part of the Hamiltonian constraint reads
$C_{\mathrm{grav}} = -6 c^2\sqrt{|p|}/\gamma^2$. It is convenient
to introduce new conjugate variables by a canonical
transformation:
\begin{align}
b:=\frac{\sqrt{\Delta}}{2}\frac{c}{\sqrt{|p|}}, \quad\quad
\nu:={\frac{4}{3\sqrt{\Delta}}}\sgn(p)|p|^\frac{3}{2},\nonumber
\end{align}
where $\Delta\, \equiv \, (2\sqrt{3}\pi\g) \, \lp^2$ is the
smallest non-zero eigenvalue of area operator in full LQG and
$\lp^2=G\hbar$.

In the kinematical Hilbert space ${\cal
H}^{\mathrm{grav}}_{\mathrm{kin}}$ of the quantum theory,
eigenstates of $\hat{\nu}$, which are labelled by real numbers
$v$, constitute an orthonormal basis as: $ \langle
v_1|v_2\rangle=\delta_{v_1,v_2} $. The fundamental operators act
on $|v\rangle$ as: $ \hat{\nu}\,\ket{v} = (8\pi\g\lp^2/3)v
\ket{v}$ and $\widehat{e^{ib}}\,\ket{v} = \ket{v+1}$. In the
improved LQC treatments \cite{improved}, the gravitational part of
the Hamiltonian operator is given in the $v$ representation by:
\begin{align}
\label{Cgrhat} \hat C_{\rm grav}
\,\ket{v}=f_+(v)\ket{v+4}+f_o(v)\ket{v}+f_-(v)\ket{v-4}
\end{align}
where
\begin{align}
f_+(v) &= \nonumber \f{27}{16} \,\sqrt{\f{8\pi}{6}} \,
\f{K\lp}{\gamma^{3/2}} \, |v + 2|\, \,\, \big| |v + 1|
- |v + 3| \big|, \\
f_-(v) &= \nonumber f_+(v - 4),\, \, \, \, \, \,  f_o(v) =  -
f_+(v) - f_-(v),
\end{align}
here $K\equiv\f{2\sqrt{2}}{3\sqrt{3\sqrt{3}}}$. As in
\cite{analy,improved}, to identify a dynamical matter field as an
internal clock, we take a massless scalar field $\phi$ with
Hamiltonian $C_{\phi} = |p|^{-\f{3}{2}}\, p_\phi^2/2$, where
$p_\phi$ denotes the momentum of $\phi$. While we choose the
standard Schr\"{o}dinger representation for $\phi$, the operator
$\widehat{1/|p|^{3/2}}$ is diagonal in the $v$ representation with
action:
\begin{align}
 \widehat{{|p|^{-\f{3}{2}}}}\ket{v} =
\left(\f{27}{16 \pi \gamma \lp^2}\right)^{3/2}K\,|v| \, \bigg| |v
+ 1|^{1/3} - |v - 1|^{1/3} \bigg|^3 \ket{v}.\nonumber
\end{align}
Collecting these results we can express the matter part of the
quantum Hamiltonian constraint as  $
 \hat{C_{\phi}} =
\frac{1}{2}\,\widehat{|p|^{-\f{3}{2}}}\, \widehat{p_\phi^2} $ and
the total constraint as $
 \hat{C}=\frac{1}{16\pi G}\hat{C}_{\rm
grav}+\hat{C}_\phi $.

To ensure that the Hamiltonian constraint operator is a viable
quantization, one needs to show that its expectation value with
respect to suitable semiclassical states  does reduce to the
classical constraint. Let us first consider the gravitational
part. Since there are uncountable basis vectors, the natural
Gaussian semiclassical states live in the algebraic dual space of
some dense set in ${\cal H}^{\rm{grav}}_{\rm{kin}}$. A
semiclassical state $(\Psi_{(b_o,\nu_o)}|$ peaked at a point
$(b_o,\nu_o)$ of the gravitational classical phase space reads:
\be (\Psi_{(b_o,\, \nu_o)}|=\sum_{v\in
\mathbb{R}}e^{-\frac{(v-v_o)^2}{2d^2}}e^{ib_o(v-v_o)}(v|,\label{coh}
\ee where $d$ is the characteristic ``width'' of the coherent
state, and $v_o$ is related to $\nu_o$ through
$\nu_o=({8\pi\gamma\ell_{\rm{P}}^2}/{3})v_o$. For practical
calculations, we use the shadow of the
 semiclassical state $(\Psi_{(b_o,\nu_o)}|$ on the regular lattice with spacing
 1, which
 is given by
 \begin{align}
 |\Psi\rangle \,=\,
\sum_{n\in\Z}\,\left[e^{-\frac{\epsilon^2}{2}(n-N)^2}\, \,e^{-i
(n-N)b_o}\right]\,\, |n\rangle ,\label{shad}
\end{align}
where $\epsilon = 1/d$ and we choose $v_o=N\in\mathbb{Z}$. Since
we consider large volumes and late times, the relative quantum
fluctuations in the volume of the universe must be very small.
Therefore we have the restrictions: $ {1}/{N}\ll\epsilon\ll1$ and
$ b_o\ll 1 $. To check that the state $(\Psi_{(b_o,\nu_o)}|$ is
indeed semiclassical, we have to calculate the expectation values
and the fluctuations of the fundamental variables. Although there
is no operator corresponding to $b$ in loop quantization, one may
define an approximating operator $
\hat{b}:=(\widehat{e^{ib}}-\widehat{e^{-ib}})/2i$ for $b\ll 1$.
Using the shadow state scheme \cite{shad}, the expectation values
and the fluctuations in the state $(\Psi_{(b_o,\nu_o)}|$ are
calculated as:
\begin{align}
&\langle{\hat{b}}\rangle =e^{-\frac{\epsilon^2}{4}}\sin
b_o\big(1+O(e^{-\frac{\pi^2}{\epsilon^2}})\big), \quad\quad
\langle \hat{v}\rangle=v_o,\label{<bv1>}\\
&(\Delta
 b)^2 =\frac{\epsilon^2}{2}\cos^2b_o+O(\epsilon^4)
 +O(e^{-\frac{\pi^2}{\epsilon^2}}),\label{<b>}\\
&(\Delta
 v)^2=\frac{1}{2\epsilon^2}(1+O(e^{-\frac{\pi^2}{\epsilon^2}})).\label{<v>}
\end{align}
We conclude that the state (\ref{coh}) is sharply peaked at
$(b_o,\nu_o)$ and the fluctuations are within specified tolerance.
The semiclassical state of matter part is given by the standard
coherent state
\begin{align}
(\Psi_{(\phi_o,p_{\phi})}|=\int
\rm{d}\phi\,e^{-\frac{(\phi-\phi_o)^2}{2\sigma^2}}e^{\f{i}{\hbar}p_\phi(\phi-\phi_o)}(\phi|\label{matterstate},
\end{align}
where $\sigma$ is the width of the Gaussian. Thus the whole
semiclassical state reads $(\Psi_{(b_o,\,
\nu_o)}|\bigotimes(\Psi_{(\phi_o,p_{\phi})}|$.

The task is to use this semiclassical state to calculate the
expectation value of the Hamiltonian operator to a certain
accuracy. In the calculation of
$\langle\hat{C}_{\mathrm{grav}}\rangle$, one gets the expression
with the absolute values, which is not analytical. To overcome the
difficulty we separate the expression into a sum of two terms: one
is analytical and hence can be calculated straightforwardly, while
the other becomes exponentially decayed out. In the calculation of
$\langle\hat{C}_{\phi}\rangle$, one has to calculate the
expectation value of the operator $\widehat{|p|^{-\frac{3}{2}}}$.
Using the Poisson resummation formula and the steepest descent
approximation, we obtain
\begin{align}\begin{split}
\langle\widehat{|p|^{-\f{3}{2}}}\rangle=&\left(\f{6}{8 \pi \gamma
\lp^2}\right)^{3/2}\,\f{K}{N}\big(1+\f{1}{2N^2\epsilon^2}+\f{5}{9N^2}+O(\f{1}{N^4\epsilon^4})\big)\\
&+O\big(e^{-N^2\epsilon^2}\big)
+O\big(e^{-\f{\pi^2}{\epsilon^2}}\big).
\end{split}\label{p32}
\end{align}
Collecting these
results we can express the expectation value of the total
Hamiltonian constraint, up to corrections of order
$1/(N^4\epsilon^4)$ and $e^{-\pi^2/\epsilon^2}$, as follows:
\begin{align}\begin{split}
\langle{
\hat{C}}\rangle=&-\frac{27}{32\pi}\sqrt{\frac{8\pi}{6}}\frac{K\ell_\mathrm{P}}{G\gamma^{3/2}}|v_o|
\big(e^{-4\epsilon^2}\sin^2(2b_o)-\frac{1}{2}(e^{-4\epsilon^2}-1)\big)
\\+&\f{1}{2}\left(\f{6}{8 \pi \gamma
\lp^2}\right)^{\frac{3}{2}}\f{K}{|v_o|}(p^2_{\phi}+\frac{\hbar}{2\sigma^2})\big(1+\f{1}{2|v_o|^2\epsilon^2}
+\f{5}{9|v_o|^2}\big).
\end{split}\label{<C>}
\end{align}
It is easy to write Eq.(\ref{<C>}) in terms of $(c,p)$ and see
that the classical constraint is reproduced up to small
corrections of order $b_o^2, \epsilon^2, 1/v_o^2\epsilon^2$ and
$\hbar/\sigma^2$. Hence, the improved Hamiltonian operator is a
viable quantization of the classical expression. Using the
expectation value (\ref{<C>}) of the Hamiltonian operator
\cite{ta}, we can further obtain an effective Hamiltonian with the
relevant quantum geometry corrections of order
$\epsilon^2,1/v^2\epsilon^2$ as
\begin{align}\begin{split}
{\cal {H}_{\mathrm{eff}}}=&-
\frac{27}{32\pi}\sqrt{\frac{8\pi}{6}}\frac{K\ell_\mathrm{P}}{G\gamma^{3/2}}\,v\,\big(\sin^2(2b)+2\epsilon^2\big)\\
&+\left(\f{8 \pi \gamma \lp^2}{6}\right)^{3/2}\f{v}{K}
\rho\big(1+\f{1}{2v^2\epsilon^2}\big) \label{Heff},\end{split}
\end{align}where
$\rho=\f{1}{2}\left(\f{6}{8 \pi \gamma
\lp^2}\right)^{3}\left(\f{K}{v}\right)^2p_{\phi}^2 $ is the
density of the matter field. Note that the neglected quantum
fluctuations of matter field would not qualitatively change the
following discussions. Then we obtain the Hamiltonian evolution
equation for $v$ by taking its Poisson bracket with $\cal
{H}_{\mathrm{eff}}$ as:
\begin{align}
\dot{v} =&3v\,\bigg(\sqrt{\f{8\pi G}{3}\rho_c}\ \bigg)\,\sin
(2b)\cos(2b) \,, \label{vdot}
\end{align}
where $ \rho_c\equiv {{\sqrt{3}}}/({16\pi^2G^2\hbar\gamma^3})$.
Note that a direct calculation shows that Eq.(\ref{vdot})
coincides with the expectation value of $[\hat{v},\
\hat{C}]/i\hbar$ in the shadow state (\ref{shad}). The vanishing
of the effective Hamiltonian constraint (\ref{Heff}) gives rise to
\begin{align}
\sin^2(2b)=\frac{\rho}{\rho_c}\left(1+\frac{1}{2v^2\epsilon^2}\right)-2\epsilon^2.\label{sin}
\end{align}
The modified Friedmann equation can then be derived from Eqs.
(\ref{vdot}) and (\ref{sin}) as:
\begin{align}
\begin{split} H^2=\f{8\pi G}{3}\rho\Big[
1-\f{\rho}{\rho_c}(1+\f{1}{v^2\epsilon^2})+\f{1}{2v^2\epsilon^2}-2\epsilon^2\f{\rho_c}{\rho}
\Big].\label{Fre}\end{split}
\end{align}
It is obvious that Eq.(\ref{Fre}) reduces to the leading order
effective Friedmann equation $H^2=({8\pi
G}/{3})\rho(1-{\rho}/{\rho_c} )$, if the terms of order
$1/(v^2\epsilon^2)$ and $\epsilon^2$ are neglected. However, as we
will see, the minus sign in front of the $\epsilon^2$ term in Eqs.
(\ref{sin}) and (\ref{Fre}) may lead to a qualitatively different
scenario from the leading order effective theory. Thus these
subleading terms cannot be neglected at will. Since all the
quantum geometry corrections come from $\langle{ \hat{C}_{\rm
grav}}\rangle$ and Eq.(\ref{p32}), the higher order corrections
are only higher order terms at the same place of $\epsilon^2$ or
$1/(v^2\epsilon^2)$. Hence, those neglected higher order
corrections cannot lead to qualitatively different effect.

\begin{figure}[!htb]
    \includegraphics[width=0.5\textwidth,angle=0]{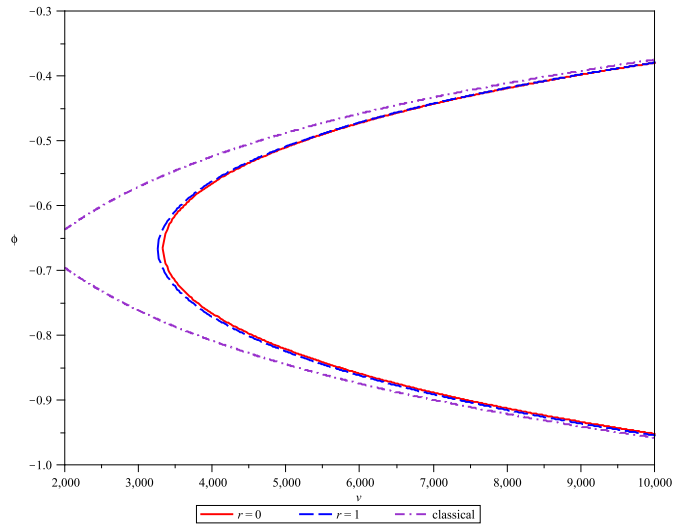}
    \caption{The effective dynamics represented by the observable
    $v|_{\phi}$ are compared to classical trajectories. In this
    simulation, the parameters were:
    $G=\hbar=1\,$, $p_{\phi}=10\,000\,$ with initial data
    $v_o=100\,000\,$,
    $\epsilon_o=0.001$.}
    \label{fig:singularity}
\end{figure}

\begin{figure}[!htb]
    \includegraphics[width=0.5\textwidth,angle=0]{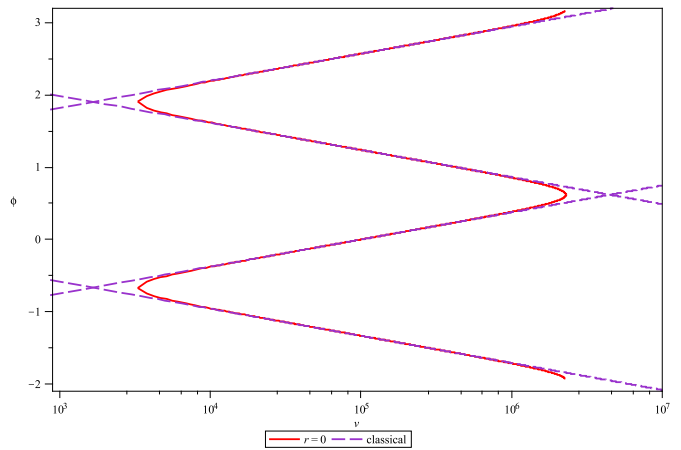}
    \caption{The cyclic model is compared with expanding and contracting classical trajectories.
    In this
    simulation, the parameters were: $G=\hbar=1$\,, $p_{\phi}=10\,000\,,\epsilon=0.001$ with initial data
    $v_o=100\,000$.}
    \label{fig:turn}
\end{figure}

It is not difficult to see that the modified Friedmann equation
(\ref{Fre}) implies significant departure from classical general
relativity, which is manifested in the bounce or collapse point
determined by $\dot{v}=0$. For a contracting universe,
Eq.(\ref{vdot}) ensures that the so-called quantum bounce will
occur when $\cos(2b)=0$. Around this point the departure of our
effective scenario from the leading order effective one is slight.
On the other hand, for an expanding universe, while the collapse
point given by $\sin(2b)=0$ will never occur in the leading order
effective scenario, it may come on stage in our scenario. The
quantum fluctuations or the Gaussian spread $\epsilon$ plays a key
role here. Thus its concrete form becomes rather relevant. One
usually sets the innocent condition that the relative spreads of
the basic conjugate variables are small for semiclassical states,
i.e., $\frac{\Delta v}{v}\sim\frac{1}{\sqrt{2}\epsilon v}\ll1$ and
$\frac{\Delta b}{b}\sim\frac{\epsilon}{\sqrt{2} b}\ll1$
\cite{bojow-r,cs}. A simple setting could be
$\epsilon=\lambda(r)v^{-r(\phi)}$, where $0\leq r(\phi)\leq1$ and
the parameter $\lambda(r)$ has to be suitably chosen for different
value of $r$. We now illustrate two extreme cases. For $r=1$, it
is easy to see that the subleading order corrections
$1/v^2\epsilon^2$ and $\epsilon^2\rho_c/\rho$ in Eq.(\ref{Fre})
are both small constant. Hence they cannot lead to significant
departure from the leading order effective scenario. While for
$r=0$, Eq.(\ref{Fre}) implies that the quantum fluctuation
correction $\epsilon^2\rho_c$ acts as a negative cosmological
constant. Thus, besides the quantum bounce when the matter density
$\rho$ increases to the Planck scale, the universe would also
undergo a collapse when $\rho$ decreases to $\rho_{\rm
coll}\approx2\epsilon^2\rho_c$. Therefore the quantum fluctuations
lead to a cyclic universe in this case. The quantum bounces in
different scenarios are compared in Fig.\ref{fig:singularity}.
Due to the quantum corrections of order $1/v^2\epsilon^2$, the
quantum bounces in $r=0$ scenario would happen at a smaller
critical density of matter than that of the $r=1$ scenario. The
trajectory of the latter almost coincides with that of the leading
order effective scenario. The cyclic universe in $r=0$ scenario is
illustrated in Fig.\ref{fig:turn}.
Since the significant departure occurs only at the large scale
limit, the asymptotic behavior of $r(\phi)$ is crucial. It is easy
to see from Eq.(\ref{sin}) that an expanding universe would
undergo the collapse and become cyclic provided $0\leq r <1$
asymptotically. Suppose that the semiclassicality of our coherent
state is maintained in the large scale limit. This means that the
quantum fluctuation $1/\epsilon$ of $v$ cannot increase as $v$
unboundedly as $v$ approaches infinity. This is another way of
saying that the quantum fluctuation $\epsilon$ of $\hat{b}$ cannot
approach zero as $b$, since otherwise the coherent state would
approach an eigenstate of $\hat{b}$ and thus lose its coherence.
In fact, the innocent condition $\frac{\Delta b}{b}\ll1$ is not
valid when $b$ approaches zero. This fact is obvious if one
recalls the standard coherent states of a harmonic oscillator,
where the fluctuation $\Delta x$ is a constant and hence
$\frac{\Delta x}{x}\ll1$ is not valid when $x$ approaches zero.
Therefore, the assumption that the semiclassicality of the model
is maintained in the large scale limit indicates a cyclic universe
driven by the quantum fluctuations. This inference is in all
probability as viewed from the parameter space of $r(\phi)$. This
is an amazing possibility that quantum gravity manifests herself
in the large scale cosmology, which has never been realized
before.

We summarize with a few remarks: (i) The main calculational result
is the effective Hamiltonian (\ref{Heff}) incorporating the next
to leading order quantum corrections, which are derived by strict
approximation schemes rather than certain simplified treatments
such as in \cite{bojow-r,non,acs}. The approximation schemes can
also be straightforwardly applied to the models with a
cosmological constant. (ii) For the scenarios of the cyclic
universe, the expectation value, infinitesimal Ehrenfest and small
fluctuation properties of the shadow coherent state (\ref{shad})
with respect to $\hat{b}$ and $\hat{v}$ are all maintained at both
the quantum bounce and collapse points. Thus the universe could be
semiclassical all the way and present its consistency. (iii)
People used to think that quantum gravity could only take effect
at small (Planck) scale. While the quantum bounce looks quite
natural, one may suspect how quantum effect can change the large
scale behavior of the universe. The intuitive picture that we
gained from this model is the following. As the universe expands
unboundedly, the matter density would become so tiny that its
effect could be comparable to that of quantum fluctuations of the
spacetime geometry. Then the Hamiltonian constraint may force the
universe to contract back. (iv) We calculate perturbatively the
effect stemming from a nonperturbative theory of quantum
cosmology. The result indicates that the subleading terms in the
effective Hamiltonian (\ref{Heff}) cannot be arbitrarily neglected
in the approximation procedure. However, our qualitative result
would not change even if corrections to higher orders are
incorporated. Our analysis does not include spatially anisotropic
or inhomogeneous quantum gravity fluctuations. Whether such
fluctuations could have a role in the inferred effect would be an
interesting question. (v) Caveats may arise from our effective
approach. Our confidence arise from the fact that the Planck scale
quantum bounce predicted by the effective Friedmann equation
(\ref{Fre}) has been confirmed by the numerical simulation in the
full quantum difference-differential system of this model
\cite{improved} and the fact that the effective Hamiltonian is
more accurate for large volumes and late times. Nevertheless, the
condition that the semiclassicality is maintained in the large
scale limit has not been confirmed. Hence further numerical and
analytic investigations to the properties of dynamical
semiclassical states in the model are desirable. It should be
noted that in a different treatment of LQC (see \cite{bojow-r}),
the dynamical coherent states and complete effective equations are
obtained, where $r(\phi)$ approaches $1$ in the large scale limit.
While that treatment leads to a quantum dynamics different from
ours, it raises a caveat to the inferred re-collapse.

We would like to thank Abhay Ashtekar, Martin Bojowald and Victor
Taveras for discussion. This work is a part of project 10675019
supported by NSFC.

\end{document}